\documentclass[twocolumn,showpacs,aps,prl,superscriptaddress]{revtex4}

\usepackage{graphicx}
\usepackage{dcolumn}
\usepackage{amsmath}
\usepackage{multirow}

\input{./pubboard/babarsym}

\def\fishlnn    {\ensuremath{\it F}\xspace}

\def\alphaeff {\ensuremath{\alpha_{\rm eff}}\xspace}

\def\Btorhopiz   {\ensuremath{\Bpm \to \rho^{\pm}\piz}\xspace}
\def\Bztopizpiz   {\ensuremath{\Bz \to \piz\piz}\xspace}

\def\de {\ensuremath{\Delta E}\xspace}
\def\cossph   {\ensuremath{|\cos{\theta_{\scriptscriptstyle S}}|\;}\xspace}

\def\mz        {\mbox{$m_{0}$}\xspace}

\newlength{\LL}
\newcommand{\etal}{{\it et al.}}

\long\def\inst#1{\par\nobreak\kern 4pt\nobreak
    {\it #1}\par\vskip 10pt plus 3pt minus 3pt}

\begin{document}

\preprint{\babar-PUB-03/028} \preprint{SLAC-PUB-10092}

\begin{flushleft}
\babar-PUB-03/028 \\
SLAC-PUB-10092\\
\end{flushleft}

\title{ {\Large \bf \boldmath Observation of the Decay \Bztopizpiz\ }
  }

%
\author{B.~Aubert}
\author{R.~Barate}
\author{D.~Boutigny}
\author{J.-M.~Gaillard}
\author{A.~Hicheur}
\author{Y.~Karyotakis}
\author{J.~P.~Lees}
\author{P.~Robbe}
\author{V.~Tisserand}
\author{A.~Zghiche}
\affiliation{Laboratoire de Physique des Particules, F-74941 Annecy-le-Vieux, France }
% ...
\collaboration{The \babar\ Collaboration}
\noaffiliation

\date{\today}

\begin{abstract}
  
  We present an observation of the decay \Bztopizpiz based on a sample
  of 124 million \BB\ pairs recorded by the \babar\ 
  detector at the \pep2\ asymmetric-energy \B\ Factory at SLAC.  We
  observe $46 \pm 13 \pm 3$ events, where the first
  error is statistical and the second is systematic, corresponding
  to a significance of $4.2$ standard deviations including systematic
  uncertainties. We measure the branching fraction $\BR(\Bztopizpiz) =
  (2.1 \pm 0.6 \pm 0.3) \times 10^{-6}$, averaged over \Bz and \Bzb decays.

\end{abstract}


\maketitle

The study of \B meson decays into charmless hadronic final states
plays an important role in the understanding of \CP violation in the
\B system. In the Standard Model, \CP violation arises from a single
complex phase in the Cabibbo-Kobayashi-Maskawa (CKM) quark-mixing matrix
$V$~\cite{ckmref}.  Measurements of the time-dependent
\CP-violating asymmetry in the \Bztopipi decay mode by the \babar\ and
Belle collaborations~\cite{sin2alpha} provide information on the angle
$\alpha \equiv \arg\left[-V_{\rm td}^{}V_{\rm tb}^{*}/V_{\rm
    ud}^{}V_{\rm ub}^{*}\right]$ of the unitarity triangle. However,
in contrast to the theoretically clean determination of the angle
$\beta$ in \Bz decays to charmonium plus neutral-kaon final
states~\cite{babarsin2beta,bellesin2beta}, the extraction of $\alpha$
in \Bztopipi is complicated by the interference of 
amplitudes with different weak phases.  The difference between \alphaeff,
derived from the measured \Bztopipi asymmetry, and $\alpha$ may be
evaluated using isospin relations between the amplitudes for the
decays $\Bz (\Bzb) \to \pip\pim$, $\Bz (\Bzb) \to \piz\piz$, and
$\Bpm\to\pipm\piz$~\cite{Isospin}.


The primary contributions to the decay \Bztopizpiz are expected to come
from the so-called color-suppressed tree and gluonic penguin
amplitudes~\cite{GHLR}.  The branching fraction for \Bztopizpiz has been
calculated in various QCD models~\cite{allTheory}.
All models use as inputs the
values of the CKM angles, typically taken from unitarity-triangle
fits. The predictions for $\BR(\Bztopizpiz)$ are in the range
$(0.3-1.1)\times10^{-6}$.

In this paper, we report the observation of the decay \Bztopizpiz
based on $(124 \pm 1) \times 10^{6}$ $\FourS\to\BB$ pairs
(on-resonance), collected with the \babar\ detector. We also use
approximately $12\invfb$ of data recorded $40\mev$ below
the \BB\ threshold (off-resonance).

\babar\ is a solenoidal detector optimized for the asymmetric-energy
beams at \pep2 and is described in detail in Ref.~\cite{babarnim}.
Charged particle (track) momenta are measured with a 5-layer
double-sided silicon vertex tracker and a 40-layer drift chamber
inside a 1.5-T superconducting solenoidal magnet.  Neutral cluster
(photon) positions and energies are measured with an electromagnetic
calorimeter (EMC) consisting of 6580 CsI(Tl) crystals.  The photon
energy resolution is $\sigma_{E}/E = \left\{2.3 / E(\gev)^{1/4}
\oplus 1.9 \right\}
\%$, and the angular resolution from the interaction point is
$\sigma_{\theta} = 3.9^{\rm o}/\sqrt{E(\gev)}$. The photon energy
scale is determined using symmetric $\piz\to\gamma\gamma$ decays.
Charged hadrons are identified with a detector of internally reflected
Cherenkov light and  ionization in the tracking detectors. The
instrumented magnetic-flux return detects neutral hadrons and
identifies muons.  High efficiency for recording \BB events in which
one \B decays with low multiplicity is achieved with a two-level
trigger with complementary tracking-based and calorimetry-based
trigger decisions.

Candidate \piz\ mesons are reconstructed as pairs of photons,
spatially separated in the EMC, with an invariant mass within $3
\sigma$ of the \piz\ mass.  The mass resolution $\sigma$ is
approximately 8~\mevcc for high-momentum \piz mesons.  Photon candidates
are required to be consistent with the expected lateral shower shape,
not to be matched to a track, and to have a minimum energy of 30 \mev.  To
reduce the background from false \piz candidates, the angle
$\theta_{\gamma}$ between the photon momentum vector in the \piz rest
frame and the \piz momentum vector in the laboratory frame is required
to satisfy $|\cos{\theta_{\gamma}}| < 0.95$.

\B meson candidates are reconstructed by combining two \piz candidates.
Two kinematic variables, used to isolate the \Bztopizpiz signal, take
advantage of the kinematic constraints of \B mesons produced at the
\FourS. The first is the beam-energy-substituted mass $\mes = \sqrt{
  (s/2 + {\bf p}_{i}\cdot{\bf p}_{B})^{2}/E_{i}^{2}- {\bf
    p}^{2}_{B}}$, where $\sqrt{s}$ is the total \epem center-of-mass (CM)
energy.  $(E_{i},{\bf p}_{i})$ is the four-momentum of the initial
\epem system and ${\bf p}_{B}$ is the \B candidate momentum, both measured in
the laboratory frame.  The second variable is $\de = E_{B} -
\sqrt{s}/2$, where $E_{B}$ is the \B candidate energy in the CM frame.
 The \de resolution for signal is approximately 80~\mev. 

The primary source of background is $\epem \to \qqbar \;(q =
u,d,s,c)$ events where a \piz from each quark jet randomly combine to
mimic a \B decay. The jet-like \qqbar background is suppressed by
requiring that the angle $\theta_{\scriptscriptstyle S}$ between the
sphericity~\cite{sphericity} axes of the \B candidate and of the
remaining tracks and photons in the event, in the CM frame, satisfies
$\cossph < 0.7$. The other source of background is \Btorhopiz
($\rho^{\pm} \to \pipm \piz$) decays
in which the charged pion is emitted nearly at rest in the \B rest frame so
that the remaining two \piz mesons are kinematically consistent with a
\Bztopizpiz decay.  Energy resolution smearing causes some \Btorhopiz
events to have  \de above the kinematic limit of $m_{B} - m_{\pi}$.
From simulation, other \B decays contribute no more than one
background event.

The number of signal \Bztopizpiz candidates is determined in an
extended, unbinned maximum-likelihood fit. The variables used in the
fit are \mes, \de, and a Fisher discriminant \fishlnn.  The \fishlnn
discriminant is a linear combination of three variables, optimized to
separate signal from \qqbar background.  The first two variables are
sums: $L_{0} = \sum_i p_i$ and $L_{2} = \sum_i p_i
\cos^{2}{\theta_i}$ where $p_{i}$ is the momentum and $\theta_{i}$ is
the angle with respect to the thrust axis of the \B candidate, both in
the CM frame, for all tracks and neutral clusters not used to
reconstruct the \B meson.  The third variable in \fishlnn is the
output of a neural network designed to separate \B events from \qqbar
background, whose inputs are information from the remaining tracks and
photons in the event.  The inputs include information about
high-momentum leptons, low-momentum leptons, charged kaons, and slow
pions (from $\Dstarp \to \Dz \pip_{\rm slow}$) in the event; these are
the same inputs used in the \B-tagging algorithm of
Ref.~\cite{babarsin2beta}.  All neural-network training and
Fisher-discriminant optimization is performed using 
simulated events.

The data are divided into two samples: a signal sample with candidates
satisfying $\mes > 5.2 \gevcc$ and $|\de|<0.2 \gev$, and a sideband
sample with candidates from on-resonance data with $\mes > 5.2 \gevcc$
and $0.2 < |\de| < 0.4\gev$ (and well outside the triangular region in
\mes and \de  populated by \Btorhopiz decays) and candidates from
off-resonance data with $\mes > 5.2 \gevcc$ and $|\de|<0.4 \gev$.  The
sideband sample contains only \qqbar background candidates and is used
in the fit to improve the statistical precision of the \fishlnn
distribution for \qqbar events.
There are 4470 events in the signal sample and 3253
events in the sideband
sample.  The reconstruction efficiency for \Bztopizpiz is $(17.7\pm2.7)\%$, and
for \Btorhopiz is $(0.8\pm 0.1)\%$, derived from simulation. The
errors are  due to a systematic uncertainty in
the efficiency for high-momentum \piz mesons to pass the selection
criteria.

For candidates in the signal sample the probabilities ${\cal
  P}_i\left(\vec{x}_j; \vec{\alpha}_i\right)$ used in the
maximum-likelihood 
fit are the product of probability density functions (PDFs)
for the variables $\vec{x}_j = \left\{\mes,\de,\fishlnn\right\}$,
given the set of parameters $\vec{\alpha}_{i}$. The likelihood function
is given by a product over all $j=1-N$ candidates and a sum over the $i =
\left\{\Bztopizpiz,\Btorhopiz,\qqbar\right\}$ hypotheses:
\begin{equation}
{\cal L}= \exp\left(-\sum_{i=1}^3 n_i\right)\,
\prod_{j=1}^N \left[\sum_{i=1}^3 n_i {\cal P}_i\left(\vec{x}_j;
\vec{\alpha}_{i}\right)
\right]\, .
\end{equation}
The coefficients $n_{i}$ are the numbers of \Bztopizpiz signal,
\Btorhopiz background, and \qqbar background events in the sample.  The
number of \Btorhopiz events is fixed in the fit to the expected
value based on the measured branching fraction $\BR(\Btorhopiz) = (11.0
\pm 2.7)\times10^{-6}$~\cite{rhopi}.  For candidates in the sideband
sample, the likelihood function includes only the PDF for the \fishlnn
variable, and only the component for \qqbar background. A simultaneous
fit to both signal sample and sideband sample data is performed.
Monte Carlo simulations are used to verify that the likelihood fit
is unbiased.

The PDFs are determined from data and simulation.  The \mes and
\de variables are correlated for both \Bztopizpiz and \Btorhopiz, so a
two-dimensional PDF derived from a smoothed, simulated distribution is
used.  The
\mes distribution for \qqbar events is modeled as a threshold
function~\cite{Argus} whose shape parameter is determined from data with
$\cossph > 0.9$. The \de distribution for \qqbar events is modeled as a
quadratic polynomial with parameters determined from data with $\mes <
5.26\gevcc$.

The PDF for the \fishlnn variable is modeled as a parametric step
function (PSF) for \Bztopizpiz, \Btorhopiz, and \qqbar events.  A PSF
is a binned distribution (as in a histogram), whose parameters are the
heights of each bin.  Since the parent distribution of \fishlnn is not
known, any functional form (such as a multiple Gaussian) assumed for
the PDF will suffer from a systematic uncertainty due to the choice of
function.  By binning the data, the PSF substantially reduces this
uncertainty.  The PSF is normalized to one, so that the number of free
parameters is the number of bins minus one.  For both \Bztopizpiz and
\Btorhopiz, the \fishlnn PSF parameters are taken from a sample of
$3.2\times10^{4}$ fully reconstructed $\Bz \to D^{(*)} n\pi
\,(n=1,2,3) $ events in data.  The \fishlnn PSF has ten bins, with bin
limits chosen so that each bin contains approximately 10\% of the $\Bz
\to D^{(*)} n\pi$ events.  Simulation is used to verify that the same
distribution can be used for both \Bztopizpiz and \Btorhopiz.  For
\qqbar background, the \fishlnn PSF parameters are free parameters in
the fit; these parameters are determined from data in both the signal
and sideband samples.

All event-selection requirements, PDF parameters, and
maximum-likelihood fit conditions were determined before
fitting the data.

\begin{figure}[!tbph]
\begin{center}
\includegraphics[width=0.49\linewidth]{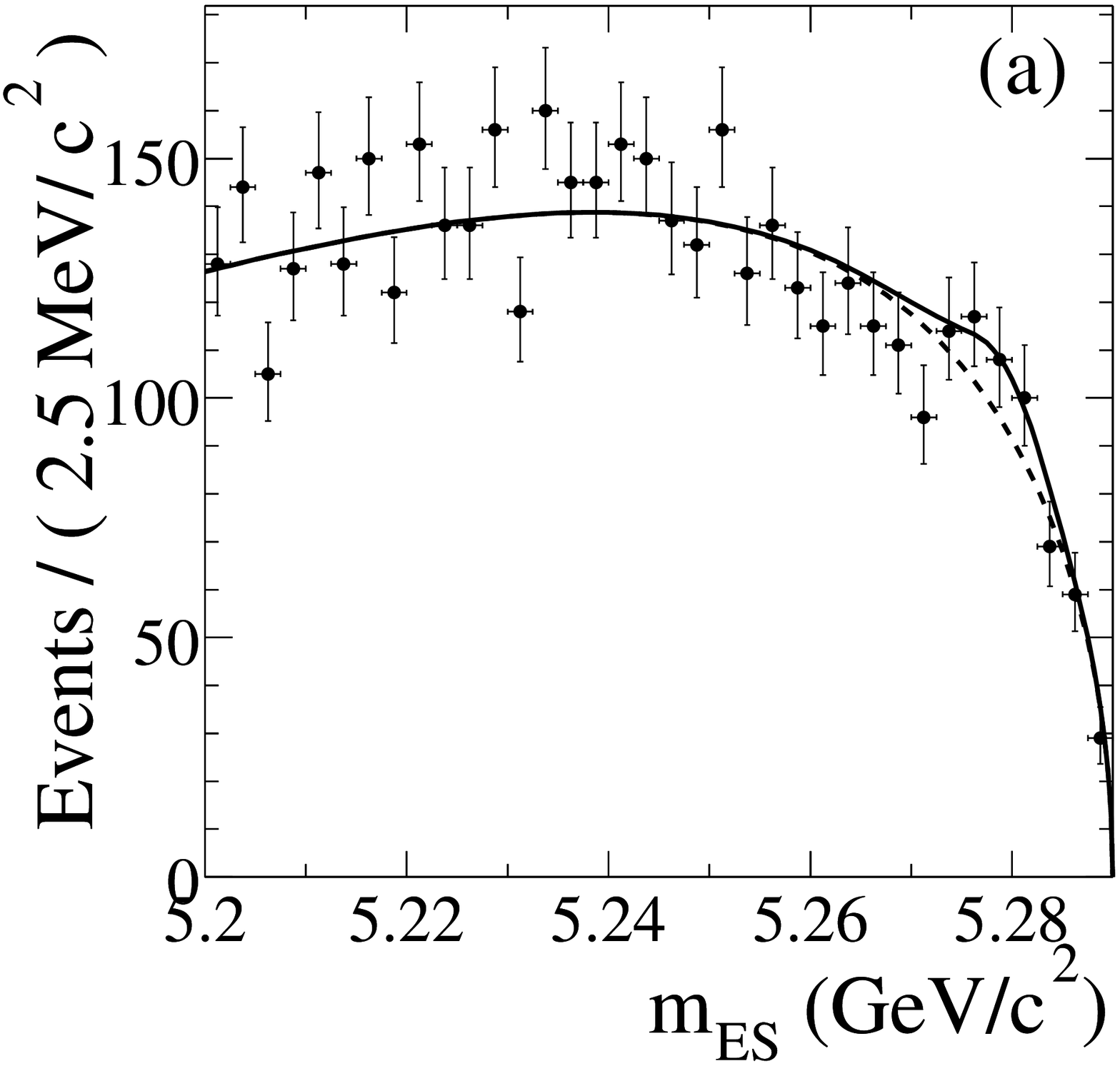}
\includegraphics[width=0.49\linewidth]{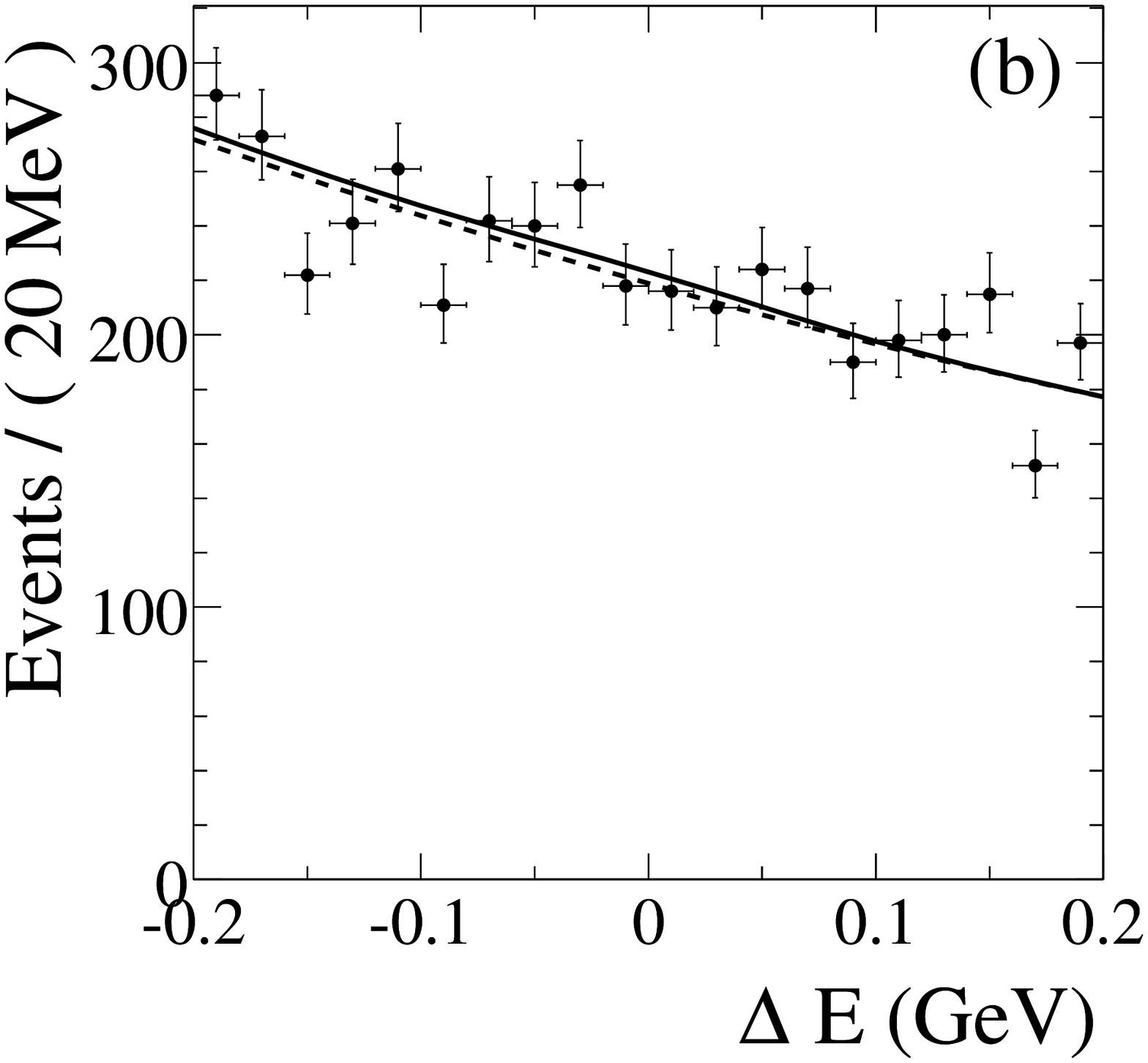}
\includegraphics[width=0.49\linewidth]{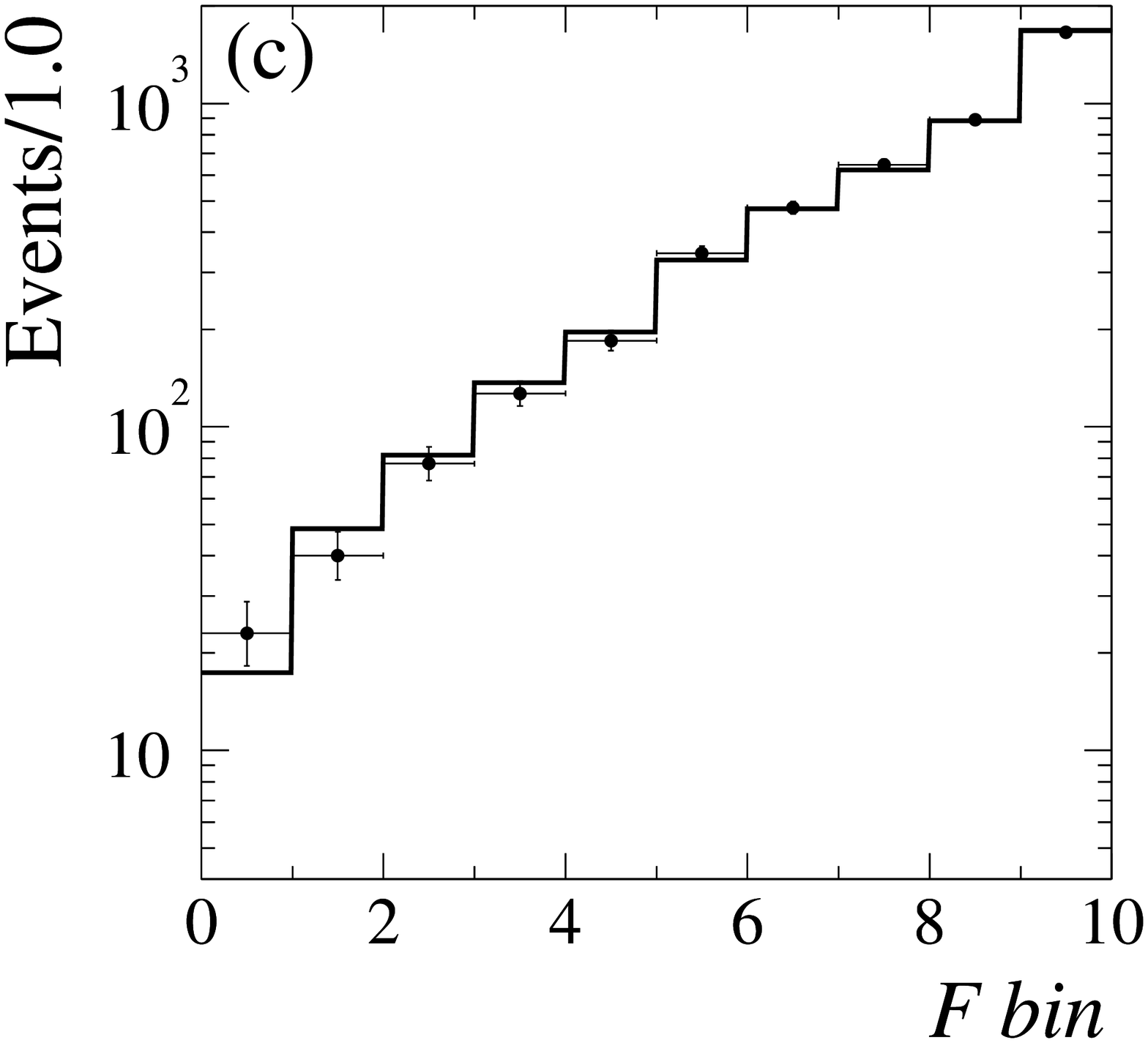}
\includegraphics[width=0.49\linewidth]{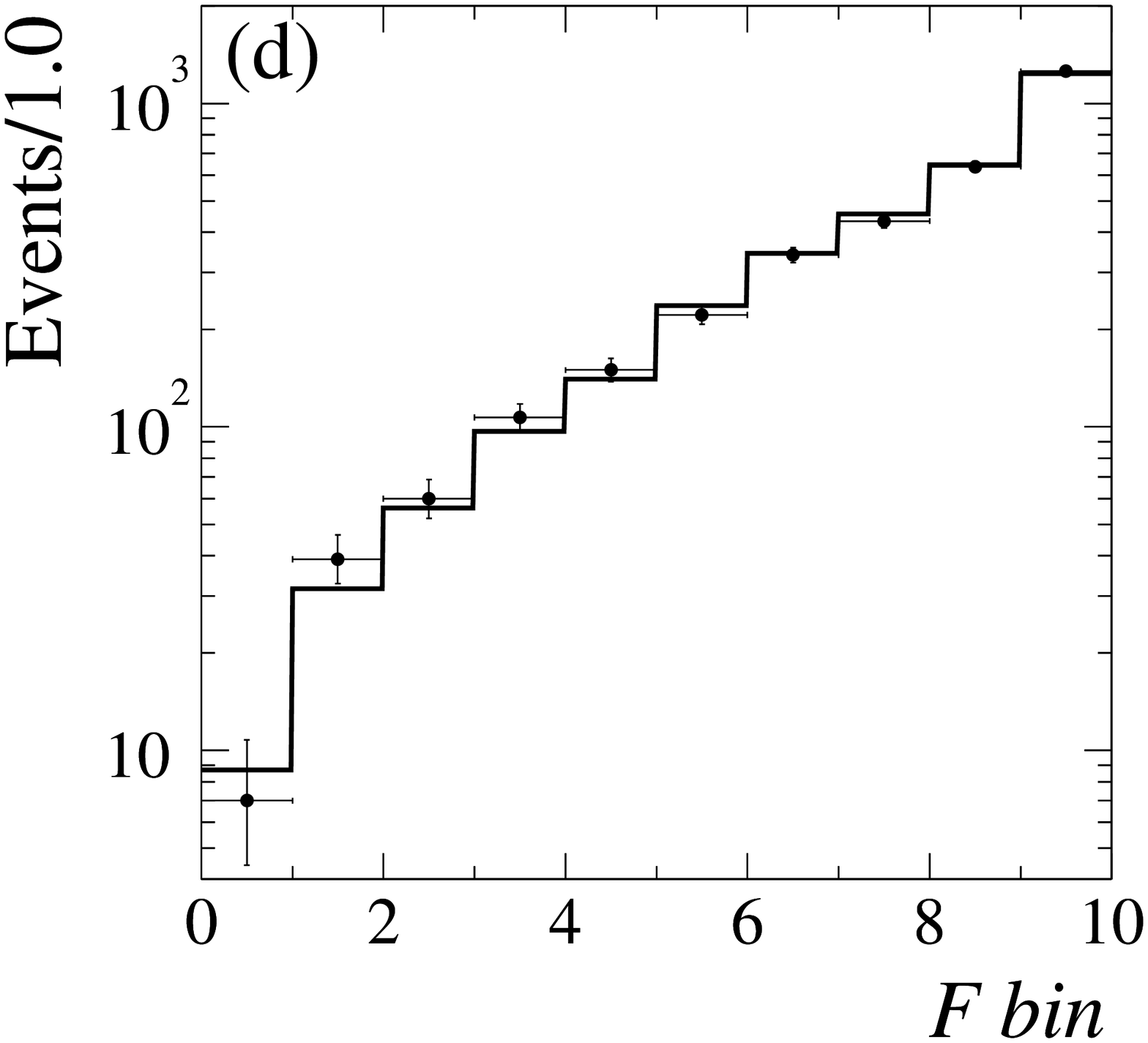}
\caption{The  distributions of a) \mes, b) \de, and c) Fisher
  discriminant \fishlnn for candidates in the signal data sample, and
  d) the \fishlnn distribution for candidates in the sideband data sample.  The
  solid lines show the PDF for signal plus background. For \mes and
  \de, the dashed lines show the PDF for the \qqbar background. The
  abscissa in c) and d) is the \fishlnn bin number, where the bins
  have been chosen so that each bin contains approximately 10\% of the
  distribution for the $\Bz \to D^{(*)} n\pi$ data sample.}
\label{fig:alldata}
\end{center}
\end{figure}

\begin{figure}[!tbph]
\begin{center}
\includegraphics[width=0.49\linewidth]{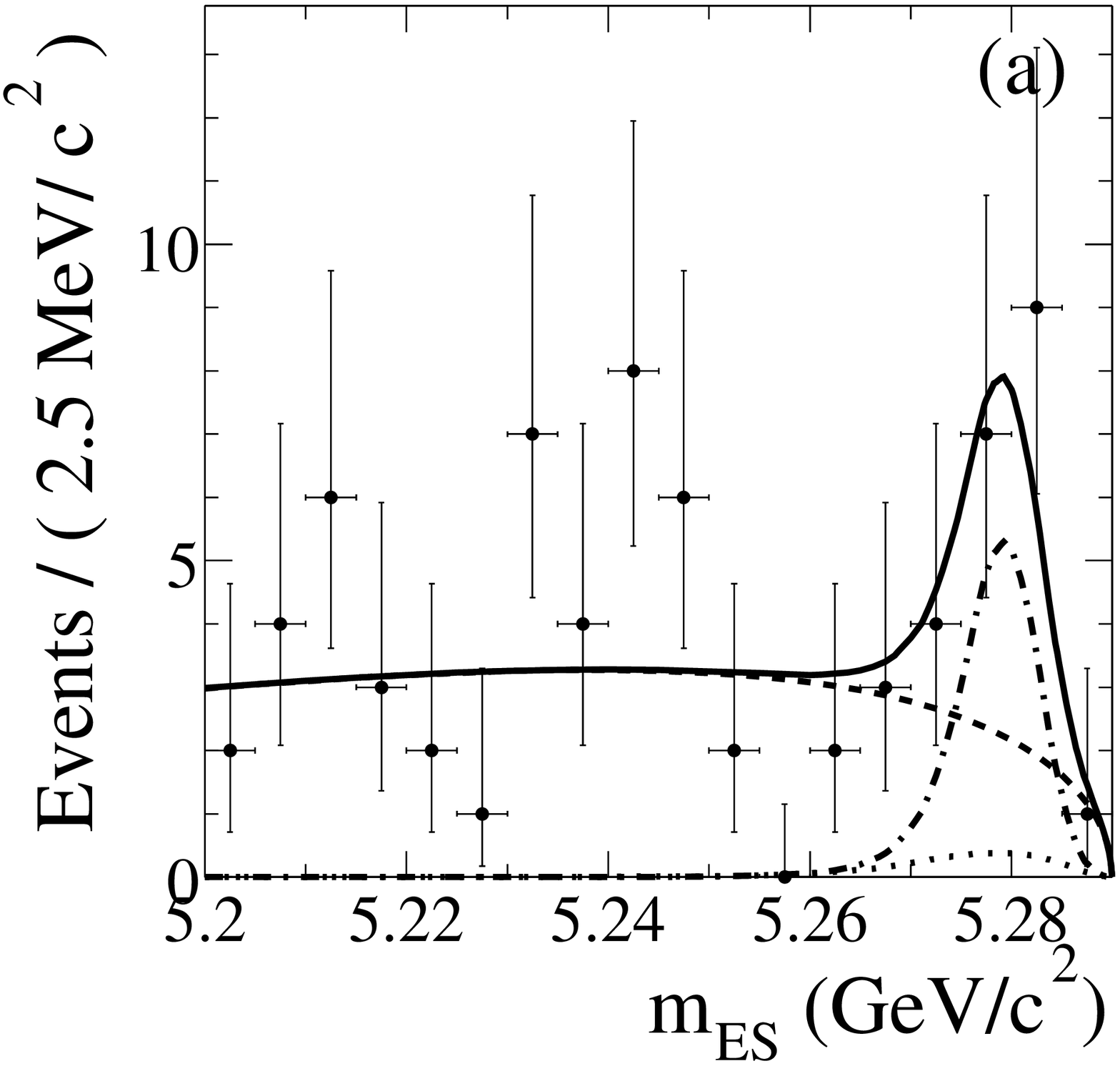}
\includegraphics[width=0.49\linewidth]{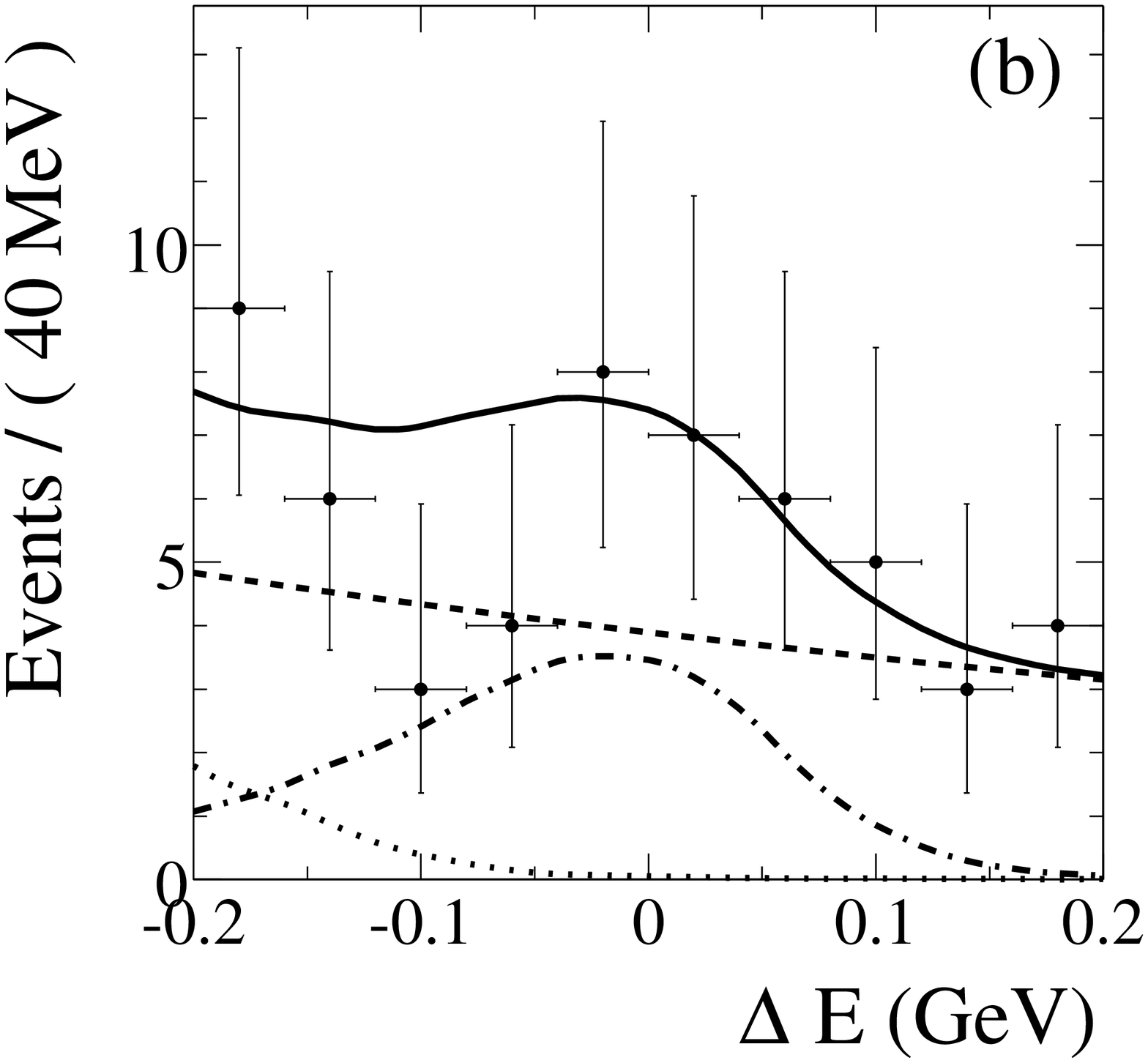}
\includegraphics[width=0.49\linewidth]{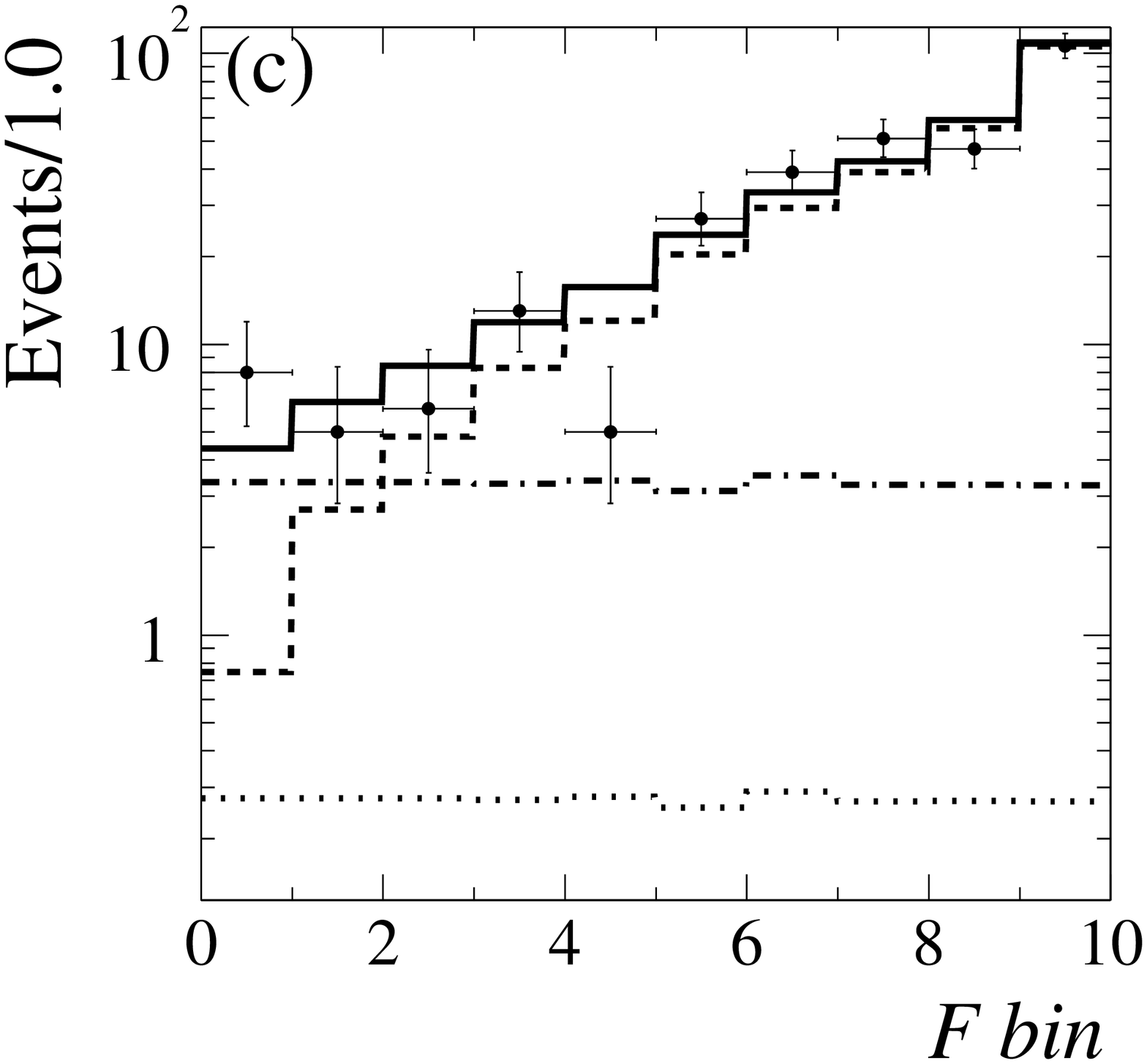}
\includegraphics[width=0.49\linewidth]{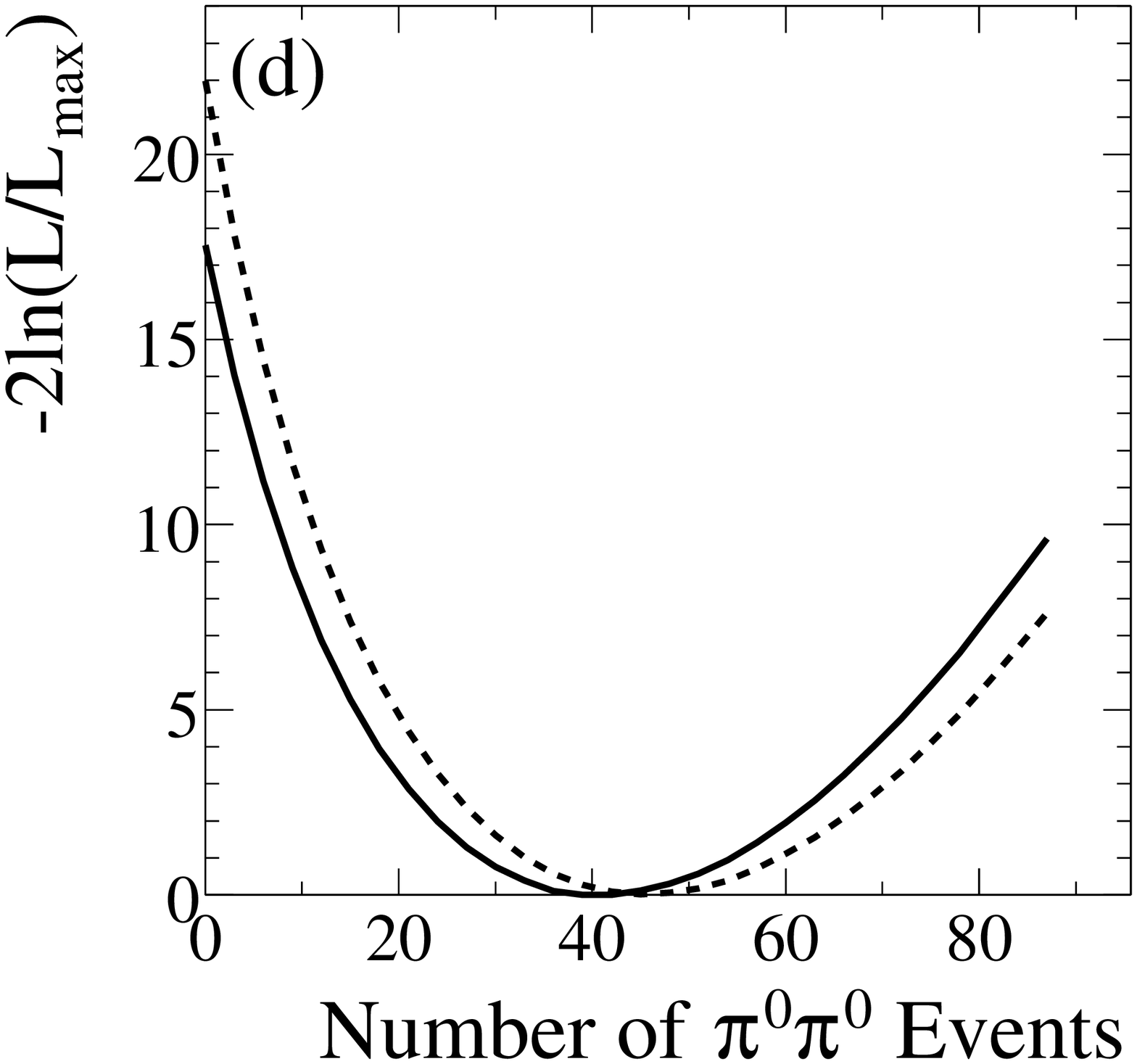}
\caption{The  distributions of a) \mes, b) \de, and c) Fisher
  discriminant \fishlnn  for
  candidates in the signal data sample that satisfy an optimized
  requirement on the signal probability, based on
  all variables except the one being plotted.  The fraction of signal
  events included in the plots is 24\%, 42\% and 74\% for \mes, \de, and
  \fishlnn, respectively. The PDF projections are shown as a dashed
  line for \qqbar background, a dotted line for \Btorhopiz, and a
  dashed-dotted line for \Bztopizpiz signal.  The solid line shows the
  sum of all PDF projections.  The PDF projections are scaled by
  the expected fraction of events passing the probability-ratio requirement.
  The ratio $-2\ln{(\mathcal{L}/\mathcal{L}_{max})}$ is shown in d)
  where the dashed line corresponds to statistical errors only and the solid
  line corresponds to statistical and systematic errors, as applied for the
  calculation of significance.}
\label{fig:projplots}
\end{center}
\end{figure}

The result of the fit is $n_{\Bztopizpiz} = 46 \pm 13$ events,
corresponding to a branching fraction of $\BR(\Bztopizpiz) = (2.1 \pm
0.6)\times 10^{-6}$.  \Bz and \Bzb decays are not separated, so the branching
fraction is measured for the average of \Bz and \Bzb. 
The \mes, \de, and \fishlnn distributions are
shown in Fig.~\ref{fig:alldata} for all data used in the fit, and in
Fig.~\ref{fig:projplots} for events that pass a requirement on the signal
probability ratio.  This requirement is optimized to maximize the ratio
$S/\sqrt{S+B}$, where $S$ is the number of signal events and $B$ is the
number of background events in the plot.  The significance of
the event yield is evaluated from the square root of the change in
$-2\ln{\cal L}$ between the nominal fit and a separate fit in which the
signal yield is fixed to zero, and is found to be $4.7\sigma$ with
statistical errors only.  

The number of signal events is stable when
the \qqbar \mes and \de PDF parameters, 
or $n_{\Btorhopiz}$, are allowed to vary in the fit.  A validation of the
maximum-likelihood fit is made by performing a simpler event-counting
analysis, based on the number of events satisfying tighter \mes, \de,
and \fishlnn requirements.  The event-counting analysis finds $13\pm
6$ events with an efficiency of $31\%$ relative to the
maximum-likelihood fit.  This agrees well with the fitted result, and
has a statistical significance of $2.7\sigma$.

This result is consistent with
our previous limit for this decay~\cite{babarpizpiz} based on
$88\times10^{6}$ \BB pairs. The data
described in Ref.~\cite{babarpizpiz} were reanalyzed with improved EMC
energy calibration and tracking alignment. More events are observed in
this data sample
after the reanalysis, consistent with the improved understanding of
the detector.

Systematic uncertainties on the event yield are evaluated by varying
the fixed parameters and refitting the data, and are summarized in
Table~\ref{tab:syst}.  The shape parameter for the threshold function
describing the \mes distribution for \qqbar events is varied to
account for the statistical error from the fit to the sample with
$\cossph>0.9$ and the extrapolation from $\cossph>0.9$ to
$\cossph<0.7$.  The \qqbar \de polynomial parameters are varied by
their statistical errors.  The number of \Btorhopiz background events
is varied according to the uncertainties on the \Btorhopiz branching
fraction and reconstruction efficiency.  Finally, the uncertainty in
the mean of the \de distribution for \Bztopizpiz is evaluated from a
study of \Btorhopiz events that have a high momentum \piz.
Extrapolating from the uncertainty in the mean \de for this sample, we
vary the mean of \de by $\pm 12~\mev$ to evaluate the systematic
uncertainty on the signal yield.  The effect of these uncertainties on
the significance of the event yield is evaluated by choosing the
variation that reduces the signal in all four systematic effects, and
then refitting the data. The significance of the signal yield after
accounting for systematic uncertainties is $4.2\sigma$.  The change in
$-2\ln{\cal L}$ as a function of the signal event yield is shown in
Fig.~\ref{fig:projplots}d.

 We observe $46 \pm 13 \pm 3$
\Bztopizpiz events with a significance of $4.2$ standard deviations
including systematic uncertainties.
We measure a branching fraction $\BR(\Bztopizpiz) = ( 2.1 \pm 0.6 \pm
0.3 )\times 10^{-6}$, where the first error is statistical and the
second is systematic.  The branching fraction is an average for \Bz
and \Bzb decays. The systematic uncertainties from PDF
variations and efficiency have been combined in quadrature. This
result is consistent with, and supersedes, our previous limit for this
decay~\cite{babarpizpiz}; it is also consistent with other prior
limits~\cite{cleobellepizpiz}. The observed \Bztopizpiz branching
fraction is larger than predicted theoretically.

\begin{table}[!htb]
\begin{center}
\caption{A summary of systematic uncertainties listed as the change in
  the fitted event yield, $\Delta n_{\Bztopizpiz}$, for different
  parameter variations.}
\label{tab:syst}
\begin{ruledtabular}
\begin{tabular}{lc}
Parameter                   & $\Delta n_{\Bztopizpiz}$   \\ \hline
\qqbar \mes shape parameter & $\pm 2.0$ \\[1mm]
\qqbar \de quadratic polynomial  & $^{+0.9}_{-1.0}$ \\[1mm]
$n_{\Btorhopiz}$            & $\pm 0.9$ \\[1mm]
$\Bztopizpiz \de$ mean      & $^{+0.6}_{-1.0}$ \\[1mm]
\end{tabular}
\end{ruledtabular}
\end{center}
\end{table}

\par

We are grateful for the excellent luminosity and machine conditions
provided by our \pep2\ colleagues, 
and for the substantial dedicated effort from
the computing organizations that support \babar.
The collaborating institutions wish to thank 
SLAC for its support and kind hospitality. 
This work is supported by
DOE
and NSF (USA),
NSERC (Canada),
IHEP (China),
CEA and
CNRS-IN2P3
(France),
BMBF and DFG
(Germany),
INFN (Italy),
FOM (The Netherlands),
NFR (Norway),
MIST (Russia), and
PPARC (United Kingdom). 
Individuals have received support from the 
A.~P.~Sloan Foundation, 
Research Corporation,
and Alexander von Humboldt Foundation.


\begin{thebibliography}{99}
\bibitem{Argus} The threshold function used for the \mes PDF is $(\mes/\mz) \sqrt{1 -
(\mes/\mz)^{2}} \, {\rm exp}\left\{-\xi\left[1 - (\mes/\mz)^{2}\right]\right\}$,
where \mz is the \mes endpoint, and $\xi$ the shape parameter.  See 
 ARGUS Collaboration, H.\ Albrecht {\em et al.}, \zpc{48}, 543 (1990).
\end{thebibliography}
\end{document}